\documentclass[aps,prb,twocolumn,amsmath,amssymb]{revtex4}

\usepackage{graphicx}
\usepackage{dcolumn}
\usepackage{bm}
\usepackage[caption=false]{subfig}
\usepackage{amssymb}
\usepackage{amsmath}
\usepackage{verbatim}
\usepackage{mathtools}
\usepackage{amsfonts}
\usepackage{physics}
\usepackage{siunitx}

\usepackage[usenames, dvipsnames]{color}
\usepackage{hyperref}
\usepackage{color}

\def\beq{\begin{equation}}
\def\eeq{\end{equation}}
\def\vk{{\bf k}}

\begin{document}

\title{Combining semi-local exchange with dynamical mean-field theory:  electronic structure and optical response of rare-earth sesquioxides}
	
	\author{James Boust$^{1}$, Anna Galler$^{1,5}$, Silke Biermann$^{1,2,3,4}$, and Leonid V. Pourovskii$^{1,2}$}

	\affiliation{
		$^1$CPHT, Ecole Polytechnique, CNRS,
		Institut polytechnique de Paris, 91128 Palaiseau, France \\
		$^2$Coll\`ege de France, 11 place Marcelin Berthelot, 75005 Paris, France\\
		$^{3}$Department of Physics, Division of Mathematical Physics, Lund University, Professorsgatan 1, 22363 Lund, Sweden\\
        $^{4}$European Theoretical Spectroscopy Facility, 91128 Palaiseau, France, Europe\\
        $^{5}$Institute of Solid State Physics, TU Wien, 1040 Vienna, Austria }
        
\begin{abstract}
	In rare-earth semiconductors, wide ligand $p$ and rare-earth 5$d$ bands coexist with localized, partially filled 4$f$ shells. A simultaneous  description for both extended  and localized states   represents a significant challenge  for first-principles theories. Here, we combine an {\it ab initio} dynamical mean-field theory approach to strong local correlations with a perturbative application of the semi-local  modified Becke-Johnson exchange potential to correct the semiconducting gap. We apply this method to  calculate the electronic structure and optical response of  the light rare-earth sesquioxides RE$_2$O$_3$ (RE= La, Ce, Pr and Nd). Our calculations correctly capture a non-trivial evolution of the optical gap in RE$_2$O$_3$ due to a progressive lowering of the 4$f$ states along the series and their multiplet structure. 2$p$ $-$ 4$f$ hybridization is found to induce a substantial upward shift for the occupied 4$f$ states occurring within the $p-d$ gap, thus reducing the magnitude of the optical gap. We show that a characteristic plateau observed in the optical conductivity in the Pr and Nd sequioxides right above their absorption edge is a fingerprint of 4$f$ states located within the $p-d$ gap.
\end{abstract}
\maketitle

\section{Introduction}

Rare-earth (RE) oxides are semiconductors with a wide range of potential applications, e.g., in electronics, optics and catalysis\cite{Gasgnier1989,Trovarelli1999,Paivasaari2005,Leskela2006,Chiu2012,Goh2017}.  
Typically, the electronic structure of such RE semiconductors comprises wide semiconducting ligand (oxygen) \textit{p} and RE 5\textit{d} bands together with  strongly correlated and localized RE 4$f$ orbitals. Their spectroscopy and optical response  are strongly affected by the location of the 4$f$ quasi-atomic states, which may appear inside -- or outside -- the $p-d$ gap. This may lead to a non-monotonous evolution of the optical gap along the lanthanide  series as, for example, observed\cite{Golubkov1995,prokofiev1996} in the  RE sesquioxides, -sulfides and -selenides RE$_2X_3$ (where RE=La...Lu, $X=$O, S, or Se). In particular, the sesquioxide series RE$_2$O$_3$ has been the subject of numerous experimental and theoretical studies. These oxides are promising high K-gate dielectrics\cite{Leskela2006,Chiu2012,Goh2017}; they also represent a prototypical case of RE semiconductors for testing various theoretical approaches to correlated  semiconductors in general.\cite{Skorodumova2001,Petit2005,Hay2006,Loschen2007,Pourovskii2007,Tomczak2007,Jacob2008,Jiang2009,Jiang2012,Amadon2012,Gillen2013,Jiang2018,El-Kelany2018}

Formulating a predictive first-principles theory for these systems is challenging, since it should properly treat both non-local and local exchange-correlation effects: the former are involved in the formation of the $p-d$ gap, while the latter stem from strong local Coulomb interactions in the RE 4$f$ shells. 
Density functional theory (DFT)\cite{Kohn1965} fails at both levels, when the Kohn-Sham band structure is used as an approximation to the electronic excitation spectrum. First, when employed in conjunction with  the standard local density (LDA) or generalized gradient approximation (GGA) exchange-correlation (XC) potentials, it systematically underestimates the $p-d$ gap\cite{Yang2012}, missing, in particular, the contribution due to the exchange-correlation derivative discontinuity\cite{Perdew1983,Sham1983}. 
Second, the KS band structure cannot capture the Mott phenomenon, usually predicting an incorrect metallic behavior for strongly correlated narrow bands, like the RE 4$f$ states. 

Two approaches have gained importance to address these limitations of standard DFT. The first one is based on hybrid XC functionals
that partially substitute the LDA or GGA exchange by the exact one. These approaches substantially improve the optical gap evolution along the RE$_2$O$_3$ series\cite{Gillen2013,El-Kelany2018} as compared to the standard LDA or GGA. However, the results depend on the amount of  exact exchange admixture $\alpha$ (though some approaches to fix $\alpha$ self-consistently have been proposed\cite{,El-Kelany2018}). Moreover, predictions of the 4$f$ states position vary strongly between different types of hybrid functionals\cite{Gillen2013}.  Another significant limitation is the inability of this approach to treat paramagnetic phases. The  sesquioxides RE$_2$O$_3$ order magnetically at low temperatures of at most several Kelvin\cite{Kolodiazhnyi2018,Lejus1976,Sala2018};  their optical properties in the paramagnetic phase are the most relevant to experiment. Though unlikely in the case of RE sesquioxides, the absorption edge in RE semiconductors may generally undergo a significant change upon the onset of magnetic order\cite{Granville2006}.

The second approach involves Green's functions many-body methods, such as the dynamical mean field theory (DMFT)\cite{Metzner1989,Georges1992} or GW\cite{Hedin1965}. The combination of DFT with DMFT, DFT+DMFT\cite{Anisimov1997_1}, successfully captures the Mott phenomenon in correlated narrow bands, even in the absence of magnetic ordering. Such an improved description of the RE 4$f$ states by DFT+DMFT resulted in a qualitatively correct shape of the optical gap evolution along the RE sesquioxide series\cite{Tomczak2007,Kolorenc2016}, reproducing also the experimental value of the gap in Ce$_2$O$_3$\cite{Tomczak2007,Pourovskii2007,Amadon2012,Kolorenc2016}. However, the underestimation of the KS $p-d$ gap is not corrected within DFT+DMFT, resulting in systematically underestimated  gaps in the Pr and Nd compounds.

It is well known that the underestimation of semiconducting gaps in the Kohn-Sham band structure of DFT can be corrected by the GW approach. Being a weak-coupling perturbative  correction to DFT, GW is not suitable for the strongly localized 4$f$ states. Hence, a combination of GW with DFT+U has been applied to RE sesquioxides\cite{Jiang2009,Jiang2012,Jiang2018}; in this method, a GW correction is applied to the self-consistent DFT+U band structure. The method predicts the  RE$_2$O$_3$  gap evolution in an overall good agreement with experiment, though a careful description for a very wide range of conduction band states is required to compute the screening\cite{Jiang2018}. Similarly to the  hybrid XC approach, GW@DFT+U of Refs.~\onlinecite{Jiang2009,Jiang2012,Jiang2018} is computationally heavy and may treat only magnetically ordered phases. In these calculations, $U$ has been treated as a parameter generally fixed at a certain value for the whole series. 

A fully consistent first-principles many-body treatment of both long-range and local correlations, e.g. by the GW+DMFT approach\cite{Biermann2003}, remains currently computationally too heavy for applications to realistic RE semiconductors. Alternatively, a many-body DMFT approach to the RE 4$f$ states has been combined  with advanced XC functionals for wide semiconducting bands. Such an implementation based on a hybrid functional was applied to Ce$_2$O$_3$\cite{Jacob2008}. The Hartree-Fock contribution to the hybrid correction was however found to lead to a large unphysical crystal field splitting\cite{Jacob2008}, inducing an orbital symmetry breaking.

In the present work we 
employ the Tran-Blaha modified Becke-Johnson (mBJ) XC potential\cite{Becke2006,Tran2009} to address the problem of the underestimation of the $p-d$ gap within DFT+DMFT. mBJ  is a semi-local exchange potential designed to mimic an exact orbitally-dependent exchange potential;  it was shown to greatly improve the quantitative values of semiconducting gaps\cite{Tran2009,Koller2011} at very moderate computational cost. Ref.~\onlinecite{Huang2016} proposed to use mBJ to
replace DFT in DFT+DMFT, and presented interesting results for the
spectral and optical properties of the band insulator YbS. However,
to the best of our knowledge, using a combination of this type 
to address the physics of a strongly correlated (or at least a 
material with a partially filled correlated shell) was never 
attempted.
In our calculations, we further calculate the values of the Coulomb repulsion $U$  from first principles. The whole framework has been previously applied to light rare-earth fluorosulfides\cite{LnSF_paper}, where we provided only a brief outline of the present methodology. Here, we provide details on the implementation and apply it to  the paramagnetic phase of light RE sesquioxides (RE=La-Nd). We find that the precision of our parameter-free scheme is comparable to GW@LDA+U and hybrid-functional approaches. Hybridization effects are found to  induce a 
significant 
systematic upward shift of the occupied 4$f$ states, hence, corrections beyond the simple quasi-atomic treatment\cite{Hubbard1963} of RE 4$f$ are necessary not only in the case of Ce, but also for Pr and Nd. 
We clarify the physical origin of this shift by analytical calculations within a simple model. 
By  simultaneously evaluating both the electronic structure and optical conductivity, we establish a connection between the position of RE 4$f$ states relative to the $p-d$ gap and the character of the optical response.

\section{Method}

\subsection{mBJ@DFT+DMFT approach}

The mBJ exchange potential can be employed either self-consistently, as, e.g., in Ref.~\onlinecite{Koller2011}, or as a perturbative correction on top of self-consistent DFT calculations\cite{Jiang2013}. One should note that  the mBJ potential is not variational, i.e., it is not derived from any XC energy functional. Correspondingly, its self-consistent application lacks a theoretical justification. Moreover, we found that such a self-consistent application of mBJ tends to  induce a symmetry breaking of the 4$f$ states, similarly to the case of hybrid functionals+DMFT\cite{Jacob2008}. This mBJ induced splitting of the 4$f$ states becomes problematic in the DFT+DMFT context, since it stems from 4$f$ exchange-correlation effects and needs to be removed  in order to avoid a double counting.  Designing such a specific double-counting term for a semilocal non-variational XC potential is highly non-trivial. However,  we found that the perturbative "one-shot" application of mBJ, which is in this case evaluated from the charge density previously converged within standard  LDA(+DMFT), induces a negligible 4$f$ splitting. Hence, following the previous implementation of mBJ@DFT+U\cite{Jiang2013}, we also employed mBJ in a perturbative fashion; we correspondingly abbreviate our approach as mBJ@DFT+DMFT.  

The  mBJ@DFT+DMFT scheme can be divided into two steps. We first carry out charge-self-consistent DFT+DMFT calculations using LDA as the XC potential, together with the Hubbard-I (HI) approximation\cite{Hubbard1963} for the DMFT impurity problem. This method is abbreviated below as DFT+HI. The quasi-atomic HI approximation has proven to reliably reproduce the  multiplet structure of localized 4$f$ states\cite{Pourovskii2009,Locht2016}; the self-consistent DFT+HI\cite{Pourovskii2007} has further shown to provide a qualitatively correct evolution of the gap along the RE$_2$O$_3$ series\cite{Tomczak2007}. We subsequently apply the perturbative mBJ corrections to the Kohn-Sham (KS) bands and recalculate the final electronic structure by performing a DMFT cycle using the HI approximation for the mBJ-corrected KS bands.   We find, however, that hybridization effects induce a significant upward shift in the position of the occupied 4$f$ states within the gap, which is missed by the HI approximation. Correspondingly, in the calculation of the final electronic structure, we effectively include the hybridization effects  through a renormalization of the on-site interaction and double-counting correction, as described in the following sections.

\subsection{Computational details}

We restricted ourselves to RE sesquioxides with the hexagonal structure P$\overline{3}$\textit{m}1 (space group 164), namely RE=La, Ce, Pr and Nd\cite{Adachi1998}, and employed experimental lattice parameters\cite{Adachi1998}.

We employed  the FP-LAPW electronic structure calculation code Wien2k\cite{Wien2k} for DFT-LDA and mBJ calculations. The spin-orbit coupling was included within the usual second variational procedure.
We used the  TRIQS\cite{triqs_main,triqs_dft_tools} implementation for DMFT and performed calculations 
for 
room temperature. Wannier orbitals representing the 4$f$ states were constructed from the KS bands inside a large energy window [-9.5eV;13.6eV] including the O 2\textit{p} states and most of the RE 5\textit{d} states.  The on-site rotationally-invariant Coulomb repulsion $\hat{H}_{U}$ 
between 4$f$ electrons was specified with the two parameters $U=F^0$ and $J_{H}$. The values of $U$ were computed by a constrained-LDA+HI (cLDA+HI) approach\cite{cDFTHI_paper} yielding $U=$ 7.5, 7.8, 8.0 eV for Ce, Pr and Nd respectively. The Hund's rule coupling $J_{H}$, which is known for 4$f$ shells to be independent of the crystalline environment, was extracted from optical spectroscopy\cite{Carnall1989}; the resulting values are $J_{H}=$ 0.73, 0.77 eV for Pr and Nd. We employed $J_{H}=$ 0.71 eV for Ce. The fully localized limit (FLL) double-counting with the nominal atomic occupancy\cite{Pourovskii2007} was used throughout. $\hat{H}_{U}$  was also included for 
the empty 4$f$ energy levels in La$_{2}$O$_{3}$. For the La $4f^0$ shell, this amounts to an upward shift due to the negative FLL DC potential by $(U-J_H)/2$, which was evaluated with the extrapolated values of $U=$ 7.3 eV and $J_{H}=$ 0.69 eV.

The optical conductivity was calculated within the Kubo linear response formalism\cite{Kotliar2006} using the implementation of Ref.~\onlinecite{triqs_dft_tools} and neglecting excitonic effects. We employed 14 000 \vk\ points in the full Brillouin zone to evaluate transport integrals.  

\subsection{Treatment of hybridization effects}

As already noted above, in our calculations we employed the self-consistent DFT+HI approach, which neglects hybridization effects in the DMFT impurity problem. To elucidate the impact of  hybridization effects, we compared  the DFT+HI spectral functions of RE$_2$O$_3$ with those computed using the numerically exact hybridization-expansion continuous-time quantum Monte Carlo method (CTQMC)\cite{Werner2006,triqs_main,Seth2016}. The use of a full rotationally-invariant interaction vertex in the presence of significant spin-orbit effects and for multi-electron 4$f$ filling is prohibitively computationally expensive for the CTQMC method at present and also prone to the sign problem\cite{Gull2011}. Hence, we were able to carry out such a CTQMC calculation with full $\hat{H}_U$ only for Ce$_2$O$_3$.  Starting from the converged DFT+HI electronic structure with the mBJ correction included as described above, we carry out a DMFT loop performing $1.4\cdot10^{7}$ Monte Carlo cycles with 200 moves per cycle. In the case of Pr$_2$O$_3$ and Nd$_2$O$_3$ we employed  a simplified density-density form for $\hat{H}_{U}$ in both the  CTQMC and HI calculation in order to investigate the impact of hybridization (see Appendix A). The off-diagonal elements of the hybridization function in the basis which diagonalizes the local 4$f$ Hamiltonian were neglected in all the CTQMC calculations.

Independently of the vertex employed,  hybridization effects are seen to induce a significant upward shift for the occupied 4$f$ peak without strongly affecting its shape and barely affect the position of the unoccupied 4$f$ states (see Appendix A).  As can be qualitatively shown by simple model calculations presented  in Appendix B,   the shift due to hybridization with occupied ligand (O 2$p$) bands primarily affects the lower Hubbard band (LHB) located just above those ligand states; the corresponding shift of  the upper Hubbard band is insignificant. 
 
Since the effect of hybridization can be effectively included through a simple shift of  the LHB, we evaluated the final electronic structure and optical response from mBJ KS bands within the HI approximation. The latter, in contrast to CTQMC, allows us to correctly take into account multiplet effects for all RE$_2$O$_3$, as well as to directly obtain the real-axis spectral function.  To include the hybridization shift, we  renormalized  $U$ and the double-counting (DC) terms by $U \rightarrow U-X$ and $\Sigma_{DC} \rightarrow \Sigma_{DC}-N_{at}X$,
with the value of $X$ chosen to align the HI and CTQMC spectral functions, yielding $X=$ 1.4, 1.6, 1.7 eV for Ce, Pr and Nd respectively. The resulting values for the normalized parameters, summarized in Table \ref{Parameters}, were then used for the final electronic spectra with HI and full rotationally-invariant Coulomb interaction.

\begin{table}[!h]
\begin{center}
\begin{ruledtabular}
	\renewcommand{\arraystretch}{1.2}
\begin{tabular}{l c c c }
      & Ce & Pr & Nd  \\
  \hline
    Calculated U & 7.5 & 7.8 & 8.0 \\
  
     Renormalized U & 6.1 & 6.2 & 6.3\\
  
    Calculated DC & 3.75 & 11.34 & 19.23 \\
  
     Renormalized DC & 2.35 & 8.14 & 14.13 \\
\end{tabular}
\end{ruledtabular}
 \caption{\label{Parameters} The value of $U$ calculated by the cLDA+HI  method and the corresponding double-counting correction $U(N_{at}-1/2)-J_H(N_{at}/2-1/2)$ for the correlated RE$_2$O$_3$ insulators. The renormalized values are reduced due to hybridization effects  (see Appendix A).}
 \end{center}
\end{table}

\section{Results}

\subsection{ Spectral properties of RE$_2$O$_3$}

\begin{figure*}[t!]
 	\begin{centering}
 		\includegraphics[width=0.99\textwidth]{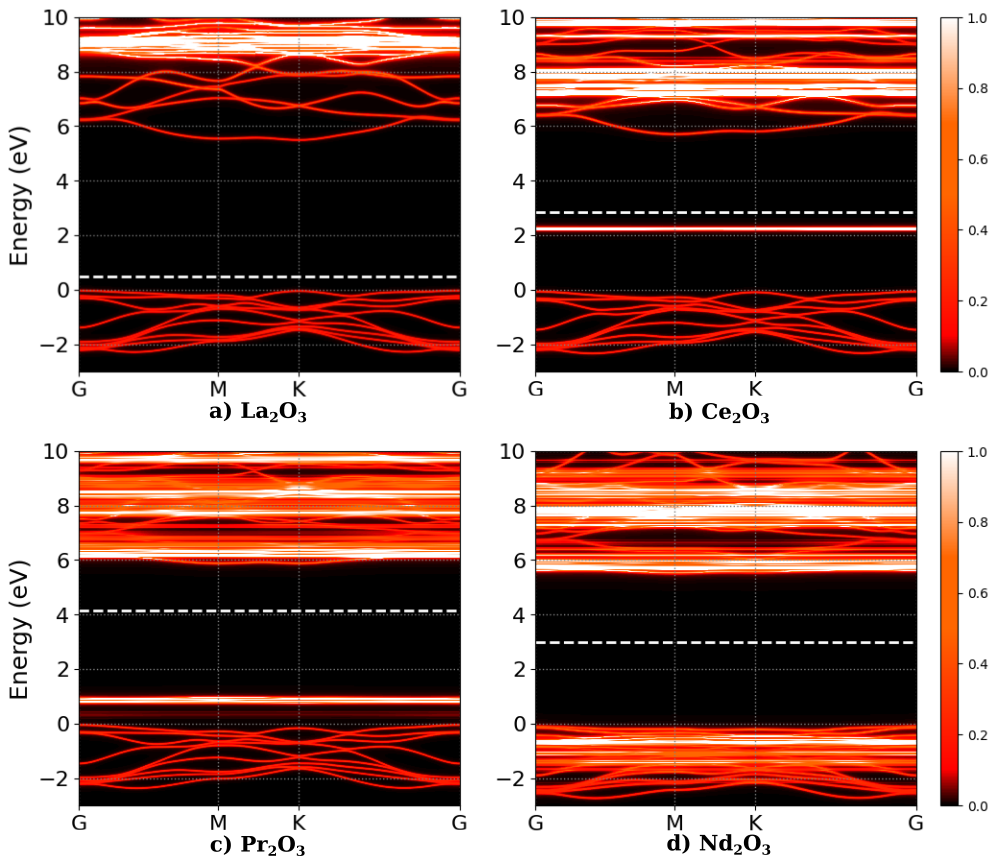}
 		\par\end{centering}
 	\caption{\vk-resolved spectral functions calculated by the mBJ@DFT+DMFT method of RE$_2$O$_3$. White color indicates a high contribution of the RE 4$f$ character. The energy is relative to the top of the O 2$p$ band. The thick dashed line is the computed chemical potential.} 
 	\label{RE2O3bands} 
 \end{figure*}

\begin{figure}[!h]
	\begin{centering}
		\includegraphics[width=0.48\textwidth]{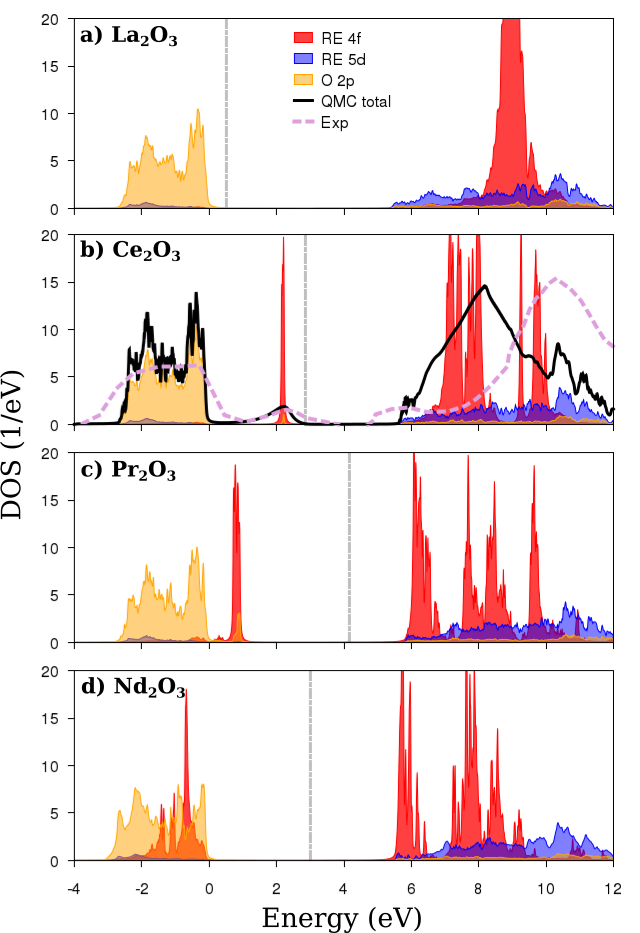}
		\par\end{centering}
	\caption{Integrated spectral functions of RE$_2$O$_3$ compounds calculated by the mBJ@DFT+DMFT method. The  energy axis zero is placed at the top of the O 2$p$ band. For Ce$_2$O$_3$, we also show the result of the CTQMC calculation with full rotationally invariant $\hat{H}_{U}$, as well as the X-ray photoemission (XPS)+inverse photoemission (BIS) measurement of Ref. \onlinecite{Allen1985}. The experimental curve was normalized to yield 18 electrons in the O 2$p$ states and 2 electrons in the Ce 4$f$ LHB. The thick dashed line is the computed chemical potential.} 
	\label{RE2O3DOS} 
\end{figure}

We first present our results for the electronic structure along the RE$_2$O$_3$ series as encoded  by the \vk-resolved and integrated spectral functions shown in  Figs.~\ref{RE2O3bands} and \ref{RE2O3DOS}, respectively.   

In La$_{2}$O$_{3}$, the empty 4$f$ states are located about 3~eV above the bottom of the La 5$d$ band, therefore,  this system is an ordinary $p-d$ band insulator  (Fig.~\ref{RE2O3bands}a). The calculated value of 5.6 eV for the optical gap between O 2$p$ and La 5$d$ is in excellent agreement with  experimental 5.4 eV\cite{prokofiev1996}. Hence, 
the  perturbative mBJ  approach successfully corrects the large underestimation of the gap within LDA, which predicts\cite{Tomczak2007} the value of  3.7~eV for the band gap in La$_2$O$_3$.

For Ce$_2$O$_3$, our calculations predict an occupied 4$f$ lower Hubbard band located within the $p-d$ gap. The absorption edge is then due to the Ce 4$f$ $-$ Ce 5$d$ optical transition as can be seen in Fig.~\ref{RE2O3bands}b.  In  Fig.~\ref{RE2O3DOS}b, our calculated integrated spectral function is compared to a X-ray photoemission (XPS)+inverse photoemission (BIS) measurement\cite{Allen1985}. The peak observed inside the $p-d$ gap agrees very well with the computed LHB position.  The width of this peak obtained by our DMFT calculations using the CTQMC method and full rotationally invariant $\hat{H}_{U}$ agrees with the experimental width of the LHB. Within the  HI  approximation, the width of LHB is strongly underestimated as expected. The computed position of the UHB is $\sim$2 eV below the one measured in Ref. \onlinecite{Allen1985}. The overall width of the UHB  in HI is controlled by multiplet splitting and is in good agreement with the CTQMC one. Separate multiplet peaks are not resolved in the latter since they are likely washed away by the analytical continuation procedure.

The \vk-resolved and integrated spectral functions of Pr$_2$O$_3$ (Figs.~\ref{RE2O3bands}c and \ref{RE2O3DOS}c, respectively) feature a narrow 4$f$ LHB located just above the top of the O 2$p$ band. The UHB has the total width of about 5 eV and is split due to multiplet effects into four main peaks. The UHB bottom is just above that of the 5$d$ conduction band, as is seen in Fig.~\ref{RE2O3bands}c.  The optical gap in Pr$_2$O$_3$ is thus between the 4$f$ LHB and the 5$d$ band.  
In Nd$_2$O$_3$, the optical transition is between the O 2$p$ states and the 4$f$ UHB (Figs.~\ref{RE2O3bands}d and \ref{RE2O3DOS}d), which is located almost precisely at the Nd 5$d$ band bottom. The magnitude of the optical gap in Nd$_2$O$_3$  is  thus predicted to be  close to that in La$_2$O$_3$.
 The LHB in Nd$_2$O$_3$ features a significant hybridization with the O-2$p$ valence band (Fig.~\ref{RE2O3DOS}d). The multiplet structure of the Nd 4$f$ UHB, with a prominent peak at the bottom of the conduction band and three more closely spaced peaks centered about 2 eV above it, is in agreement with previous calculations \cite{Locht2016,LnSF_paper} and measurements \cite{Lang1981,pauwels_thesis} for other Nd systems. To our awareness, no high-resolution photoemission measurements have been reported for Pr$_{2}$O$_3$ and Nd$_{2}$O$_{3}$.

 Our values for  the optical gap extracted from the calculated electronic structure  are listed in  Table~\ref{Gaps} together with the results of previous calculations and experimental measurements\cite{prokofiev1996}.
 Overall, our predicted gap values are in good agreement with experiment. In particular, in contrast to the previous LDA+DMFT study\cite{Tomczak2007}, both band (La$_2$O$_3$) and correlated insulators are well described. Looking at the overall picture along the series, the evolution of the optical gap is due to a progressive downwards shift of the Hubbard bands and their changing width due to multiplet effects. The semiconducting $p-d$ gap remains, to a good precision, constant along the series.
 
 However, one may also notice some systematic overestimation of the gaps  by our calculations. This overestimation can be partially due to lifetime broadening of the 4$f$ states, which is neglected within the HI approximation. For example, the HI gap in Ce$_2$O$_3$ is reduced to 3~eV with the full CTQMC treatment  (cf. Fig.~\ref{RE2O3DOS}b). Moreover, in the case of Ce$_2$O$_3$, there is only a single optical gap measurement reported in the literature and the conduction band onsets of XPS+BIS\cite{Allen1985} and XPS+XAS\cite{Mullins1998} spectra do not agree: the latter does not exhibit the shoulder at $\sim$5~eV on Fig. \ref{RE2O3DOS}b, resulting in a larger $f - p$ gap in agreement with our calculations. The actual optical gap of Ce$_2$O$_3$ might therefore be larger than 2.4~eV, as already argued in Ref. \onlinecite{Jiang2018}.

\begin{table}[!h]
\begin{center}
		\begin{ruledtabular}
		\renewcommand{\arraystretch}{1.2}
\begin{tabular} {l | c c c c }
      & La & Ce & Pr & Nd  \\
  \hline
     Exp.\textsuperscript{\cite{prokofiev1996}} & 5.4 & 2.4 & 3.9 & 4.7 \\
     HSE03\textsuperscript{\cite{Gillen2013}} & 4.5 & 2.37 & 3.5 & 4.32 \\
   sX-LDA\textsuperscript{\cite{Gillen2013}} & 5.5 & 1.75 & 3.8 & 4.65 \\
   GW@LDA+U & 4.95\cite{Jiang2012}, 5.8\cite{Jiang2018} & 1.5\cite{Jiang2012}, 3.57\cite{Jiang2018} & 2.86\cite{Jiang2012} & 4.5\cite{Jiang2012} \\
   LDA+DMFT\textsuperscript{\cite{Tomczak2007}} & 3.7 & 2.1 & 3.8 & 4.1 \\
    mBJ@DFT+DMFT & 5.6 & 3.3 & 4.7 & 5.6 \\
\end{tabular}
\end{ruledtabular}
 \caption{\label{Gaps}Calculated optical gaps of the RE sesquioxides compared to experiment as well as to previous theories. 
 	The experimental values given here are the most reliable according to Ref. \onlinecite{prokofiev1996} as they were measured on single crystals.
 }
 \end{center}
\end{table}

\subsection{Optical conductivity}

\begin{figure*}
	\begin{centering}
		\includegraphics[width=0.99\textwidth]{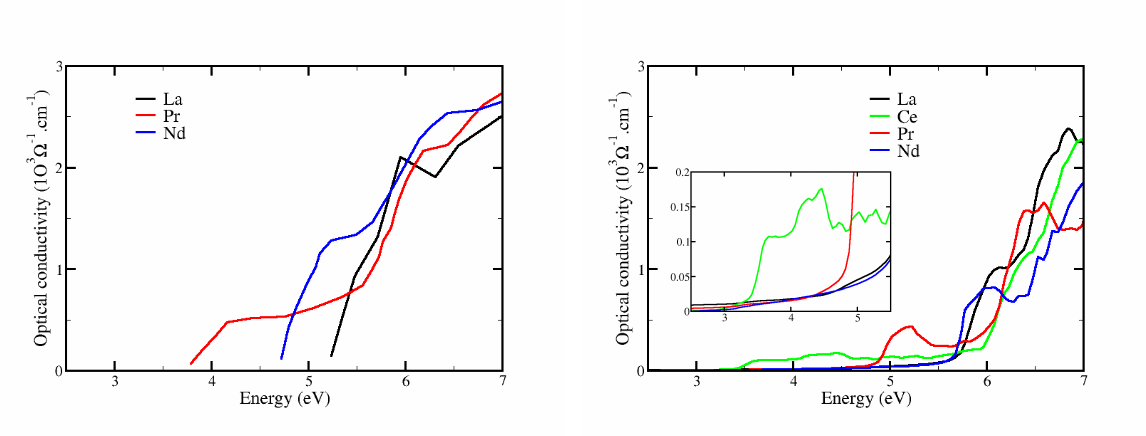}
		\par\end{centering}
	\caption{Measured\cite{Kimura2000} (left panel) and calculated in this work (right panel) optical conductivity of $R_{2}$O$_{3}$. Inset: zoom in the low energy region.} 
	\label{OC} 
\end{figure*}

Our calculated  optical conductivity is shown in Fig.~\ref{OC} together with the corresponding  experimental data of Kimura {\it et al.}\cite{Kimura2000}. 

In the correlated oxides  Ce$_{2}$O$_{3}$, Pr$_{2}$O$_{3}$ and Nd$_{2}$O$_{3}$, the theoretical optical conductivity exhibits a characteristic shape. Its onset differs significantly between the three systems, in agreement with their calculated  optical gaps  (Table~\ref{Gaps}). However, all three
optical conductivities are seen to merge at roughly 6.3 eV. At this point the conductivity is dominated by $p-d$ transitions; therefore, the approximately constant $p-d$ gap along the series results in a similar magnitude of  the corresponding contribution to the conductivity.  The evolution of the optical conductivity before this turning point is a signature of 4$f$ states involved into the absorption edge.  Correspondingly, the theoretical optical conductivity for Pr$_{2}$O$_{3}$ and Nd$_{2}$O$_{3}$  features a shallow peak 
due to the presence of the \textit{f} states inside the $p-d$ gap (or at its edges, as in Nd$_2$O$_3$).   
Schematically,  the optical conductivity increases until the whole spectral weight of the \textit{f} states within the $p-d$ gap is involved in the optical transitions; then a saturation occurs until $p-d$ transitions start contributing significantly. We observe the same kind of behavior in Ce$_{2}$O$_{3}$ though the magnitude of the initial plateau is significantly weaker due to a small spectral weight of  Ce 4\textit{f} states within the gap.  In La$_2$O$_3$, there are no 4$f$ states in the gap, hence, the initial shallow peak is missed.

Our  theoretical  picture is in a good qualitative agreement with the measured optical conductivity in La$_2$O$_3$, Pr$_2$O$_3$ and Nd$_2$O$_3$: an initial  plateau followed by a rapid increase of conductivity at the $p-d$ edge is also observed in the experimental curves of correlated insulators (Fig.~\ref{OC}, left panel), but is absent in La$_2$O$_3$. The shift between experimental and theoretical onsets of the conductivity is  consistent with our general $\sim$0.8 eV overestimation of the optical gaps in the correlated insulators.
 No optical conductivity measurement for Ce$_2$O$_3$ has been, to our awareness, reported in the literature. 
 
 We note that previous DFT+U\cite{Singh2006} and DFT+DMFT\cite{Tomczak2007}  works could not explain the occurrence of the initial plateaus  in Pr$_2$O$_3$ and Nd$_2$O$_3$. Due to their systematic underestimation of the $p-d$ gap, these calculations predicted no 4$f$ states within it. Furthermore, at higher energies, these previous works yielded theoretical values of the conductivity which were larger than the experimental ones by a factor $\sim$2; although not shown on Fig. \ref{OC}, our results exhibit the same discrepancy.

\section{Conclusion}

In this work, we proposed an {\it ab initio} approach to the electronic structure and optical properties of rare-earth semiconductors with coexisting semiconducting bands and correlated 4$f$ states.  Our methodology  combines a DMFT treatment of  strong local correlations in partially filled 4$f$ shells with an improved treatment of $p-d$ semiconducting gaps by the semi-local modified Becke-Johnson (mBJ) exchange potential. The mBJ correction is implemented in a perturbative way on top of the fully self-consistent DFT+DMFT electronic structure. In contrast to previous advanced theoretical approaches to such correlated 4$f$ electron semiconductors \cite{Jiang2012,Gillen2013}, our method is applicable to the paramagnetic phases of correlated semiconductors. Given the typical low magnetic ordering temperatures, those paramagnetic phases are more readily accessible experimentally than
the ordered ground states.

 Applying the present methodology to the light rare-earth sesquioxide series RE$_2$O$_3$ (RE=La-Nd), we obtain a qualitatively correct evolution for the optical gaps and optical conductivities along this series using {\it ab initio} values for the on-site Coulomb repulsion $U$ in the 4$f$ shells.  The precision of our scheme for the optical gaps is comparable to previous advanced  {\it ab initio} methods\cite{Jiang2012,Gillen2013}. The overall evolution of the electronic spectra and optical conductivity originates from a progressive downward shift of the 4$f$ Hubbard bands along the series as well as in the multiplet structure of those bands. Though the 4$f$ shells are usually assumed to be quasi-atomic, our study reveals a significant correction to the position of occupied 4$f$ states due to hybridization effects. Our calculations explain a characteristic shape of the measured experimental optical conductivity in the Pr and Nd sesquioxides, in which a plateau right above the  absorption edge is shown to be induced by 4$f$ states located inside the semiconducting $p-d$ gap. Therefore, the present approach is a promising tool for predicting and analyzing the optical response in correlated 4$f$ semiconductors, as already shown by its initial application to the light rare-earth fluorosulfides series\cite{LnSF_paper}.\\

\section*{Acknowledgments}
We are grateful to the computer team at
CPHT for support.

\appendix
 \section{Spectral function of RE$_2$O$_3$: impact of hybridization effects}

\begin{figure}[!h]
	\begin{centering}
		\includegraphics[width=0.48\textwidth]{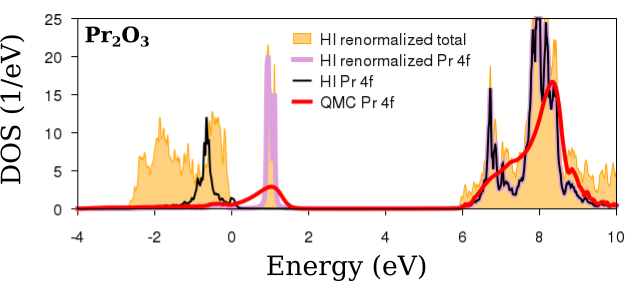}
		\par\end{centering}
	\caption{Integrated spectral function of Pr$_{2}$O$_{3}$ computed within the density-density approximation for the Coulomb interaction, with different flavors of the mBJ@DFT+DMFT approach: HI, CTQMC and HI with renormalized $U$ and DC.} 
	\label{HIvsQMC} 
\end{figure}

Fig.~\ref{HIvsQMC} shows the integrated spectral function of Pr$_{2}$O$_{3}$ computed with different approaches to the quantum impurity problem in the final DMFT run  of the mBJ@DFT+DMFT scheme. The density-density approximation for the Coulomb interaction was employed in all these benchmarks. Our CTQMC calculations within the density-density approximation were performed with $1.6\cdot 10^{8}$ Monte Carlo cycles and 400 moves per cycle. Analytical continuation of the CTQMC self-energies, which are calculated on the imaginary-frequency Matsubara grid, was performed using the maximum entropy method implemented in the TRIQS library\cite{maxent}.

The comparison between Hubbard-I and CTQMC, both with \textit{ab initio} values of $U$ and DC, illustrates the hybridization-induced shift of the 4$f$ LHB, with the UHB being unaffected. This shift can effectively be taken into account at the level of HI by renormalizing  $U$  and  correcting correspondingly the DC term. By  setting $U \rightarrow U-X$ and $\Sigma_{DC} \rightarrow \Sigma_{DC}-N_{at}X$ one may align the HI spectra with the CTQMC one, as illustrated on Fig.~\ref{HIvsQMC}. This unambiguous renormalization scheme yields $X=1.4$ eV for Pr$_{2}$O$_{3}$. The same procedure was applied to the other correlated insulators and the renormalized parameters are summarized in Table \ref{Parameters}.

\section{Perturbative treatment of the 4$f$ hybridization shift}

The hybridization shift of the LHB  observed in the RE$_2$O$_3$ spectral function (e.~g., Pr$_2$O$_3$, Fig.~\ref{HIvsQMC}) can be qualitatively understood within a simplified model using a second-order perturbation-theory treatment of the RE 4$f$ - O 2$p$ hybridization. 

We consider the 4$f$ quasi-atomic Hamiltonian within the density-density approximation for the interaction vertex: 
\begin{equation}
\label{H}
    \hat{H}_{4f} =\sum_{i}(\epsilon_{i}-\Sigma_{DC})f^{\dagger}_{i}f_{i}+\frac{1}{2}U \sum_{i\neq j}\hat{n}_{i}\hat{n}_{j}\\
\end{equation}
where $\epsilon_{i}$ is the KS energy level of the orbital $i$ of the 4$f$ shell, $U$ is the Coulomb parameter and $\Sigma_{DC}=U(N_{at}-1/2)$ is the FLL double counting correction with the nominal atomic occupancy $N_{at}$. The $\epsilon_{i}$ are close to the Fermi level, which we set as the zero energy; we will therefore take $\epsilon_{i}\approx 0$ eV.

We then add the hybridization of the 4$f$ shell  with 2$p$ shells of the three  O ions that are the nearest neighbors of the RE site. Hence, we include $3\times 6$ occupied 2$p$ levels at an average energy $\epsilon_{p}$:
\begin{equation}
\label{fH}
    \hat{H} =\hat{H}_{4f}+\epsilon_{p}\sum_{j}c^{\dagger}_{j}c_{j}-\sum_{ij}t(f^{\dagger}_{i}c_{j}+c^{\dagger}_{j}f_{i})
\end{equation}
where $t$ is the hybridization matrix element.

Diagonalizing the full Hamiltonian (\ref{fH}) with finite $t$ is analogous to an exact approach taking into account the hybridization of 4$f$ orbitals with O 2$p$ states; assuming $t=0$ is analogous to the HI approximation.

We first consider the limit $t=0$ of this simplified model. The eigenstates with fully occupied $p$ states can be labelled by the number $N$ of electrons in the 4$f$ shell, the ground-state occupancy being $N_{at}$. We can compute the following eigenenergies:
\begin{align*}
  &  E_{N_{at}}  = 18\epsilon_{p}-U\frac{N_{at}^{2}}{2}\\
   & E_{N_{at}-1}= E_{N_{at}+1} =18\epsilon_{p}-U\frac{N_{at}^{2}-1}{2}.
\end{align*}
The LHB and UHB energies are given by:
\begin{align*}
    & E_{LHB} =E_{N_{at}}-E_{N_{at}-1}= -\frac{U}{2}\\
    & E_{UHB} =E_{N_{at}+1}-E_{N_{at}}=\frac{U}{2}.
\end{align*}

We now consider the effect of small finite hybridization $t$ within the second-order perturbation theory on these different energy levels -- the first order has no contribution. As the O 2$p$ states are occupied, the only possible processes are the hopping of their 18 electrons to the $14-N$ 4$f$ empty states and back, hence
\begin{align*}
    & E_{N_{at}}(t)\approx E_{N_{at}} -\frac{18(14-N_{at})t^{2}}{-\epsilon_{p}+U/2}\\
   & E_{N_{at}-1}(t)\approx E_{N_{at}-1}-\frac{18(14-N_{at}+1)t^{2}}{-\epsilon_{p}-U/2}\\
&    E_{N_{at}+1}(t)\approx E_{N_{at}+1} -\frac{18(14-N_{at}-1)t^{2}}{-\epsilon_{p}+3U/2}
\end{align*}
and therefore 
\begin{align*}
    & E_{LHB}(t) \approx E_{LHB}-\frac{18(14-N_{at})t^{2}}{-\epsilon_{p}+U/2}+\frac{18(14-N_{at}+1)t^{2}}{-\epsilon_{p}-U/2}\\
    & E_{UHB}(t) \approx E_{UHB}-\frac{18(14-N_{at}-1)t^{2}}{-\epsilon_{p}+3U/2}+\frac{18(14-N_{at})t^{2}}{-\epsilon_{p}+U/2}.
\end{align*}
In the RE$_{2}$O$_{3}$ systems studied in this work, we typically have $\epsilon_{p}\sim -5$ eV and $U\sim 8$ eV, which gives $1/(-\epsilon_{p}+U/2)\sim 0.1$ eV$^{-1}$, $1/(-\epsilon_{p}+3U/2)\sim 0.06$ eV$^{-1}$ and $1/(-\epsilon_{p}-U/2)\sim 1$ eV$^{-1}$. Therefore, as a first approximation, hybridization induces a shift of the LHB  towards higher energies by lowering the energy of the $N=N_{at}-1$ sector, but does not affect the UHB:
\begin{align}
\label{shift}
    E_{LHB}(t)& \approx E_{LHB}+18(14-N_{at}+1)\frac{t^{2}}{-\epsilon_{p}-U/2}\\
    E_{UHB}(t)& \approx E_{UHB} \nonumber
\end{align}
The value of $t^{2}$ in the RE sesquioxides was estimated by the average over all $p$ and $f$ orbitals $t^{2}=\frac{1}{18\times14}\sum_{i\in [1:14], j \in [1:18]}|V_{ij}|^{2}$ where $V_{ij}$ is the hopping between $f$ orbital $i$ and $p$ orbital $j$. The matrix $V$ was computed within the mBJ@DFT+DMFT approach by $V=\sum_{\bold{k} \in BZ}P_{\bold{k}}^{f}H_{\bold{k}}^{KS}(P_\bold{k}^{p}){\dagger}$ where $H_{\bold{k}}^{KS}$ is the KS Hamiltonian and $P_{\bold{k}}^{f(p)}$ the projection to RE 4$f$ (O 2$p$) Wannier orbitals. This procedure yielded $t^{2}=0.01$ eV in Ce$_{2}$O$_{3}$, hence a LHB shift of $\sim 2.5$ eV from (\ref{shift}) (for $N_{at}=1$).

The results of this simplified model are perfectly consistent with the hybridization shift of the LHB computed by the CTQMC approach (illustrated on Fig.~\ref{HIvsQMC} in Appendix A) which can therefore be explained by the hopping of O 2$p$ electrons to the 4$f$ states and back.


\end{document}